\documentclass{article}
\usepackage{amsmath,latexsym,amssymb,amsthm}
\usepackage{epsfig,graphicx,subfigure}
\usepackage[overload]{textcase}
\usepackage{authblk}
\allowdisplaybreaks[1]
\normalsize
\begin{document}
\title{A sorting algorithm for complex eigenvalues}
\author[]{Usha Srinivasan \thanks{usha.s@nal.res.in}} 
\author[]{ Rangachari Kidambi \thanks{kidambi@nal.res.in}}
\affil{Computational \& Theoretical Fluid Dynamics Division, 
National Aerospace Laboratories, Bangalore 560017, India.}

\maketitle
\begin{abstract}
We present SPEC-RE, a new algorithm to sort complex eigenvalues, generated
as the solutions to algebraic equations, whose coefficients are analytic 
functions of one or 
many, possibly complex parameters. The fact that the eigenvalues are analytic
functions of the parameters, with atmost algebraic singularities, is used in
formulating the algorithm. Several examples are presented to demonstrate the
efficacy of the method; simple examples where other methods fail are also 
presented. The algorithm is likely to be useful in several problems of physics  
and engineering that require identification and sorting of eigenmodes.
\end{abstract}
\section{Introduction} \label{sec:introduction}
Linear stability analysis is ubiquitous in physics and engineering and basic
to such analysis is computation of spectral maps\footnote{The terms `modal
maps,' `eigenvalue trajectories' and `eigenpaths' will also be used to denote 
these maps.}i.e 
identifying eigenmodes as a function of the system parameters.   

There are two broad approaches to producing such eigenmodes. The first one, 
which considers the parameter variations as perturbations and then computes the
new eigenvalues as responses to these perturbations, has a long history dating
back to Jacobi in the middle of the nineteenth century. Since then, a variety 
of methods like power methods and invariant subspace methods have been 
developed. A nice recent review
of some of these methods is \cite{Golub-2000}. Starting with the original 
system eigenvalues, the various algorithms provide means to compute first-order
eigenvalue sensitivities which are then used in a rapidly convergent iterative 
method to produce new eigenvalues corresponding to the parameter change.
Some applications of this approach are \cite{Beyn-2009}, \cite{Eldred-1995}
and \cite{Wagner-2003}. We call this the `tracing' approach. Note that the
eigenmodes are produced automatically in course of the tracing. The second
approach is to solve the so-called dispersion relation (DR) for a range of
parameter values and then to sort the already determined eigenvalues, at
different values of the relevant parameter. Though this is mathematically 
simpler, it seems to have been explored relatively less; some publications, in 
fluid mechanics (\cite{Koch-1986}, \cite{Suslov-2006}) and electromagnetics 
(\cite{Capek-2011} ) use this approach. We call this the `sorting' approach.
The `sorting' approach is the approach to use if one is interested in 
identifying a few (typically 10-20) eigenmodes over a large range of parameter 
values for moderate sized matrix systems (typically $N < 5000$) whereas the 
`tracing' approach 
has to be the preferred one if the goal is to generate very few (typically 1-2) 
eigenmodes for a limited range of parameter values, but for large matrix 
systems (typically $N > 5000$).

For real problems, the DR is rarely known in closed form; it exists as the
outcome of a matrix eigenvalue problem and is only numerically known (NDR).
Solution of such NDR is not really a problem; indeed fast and accurate numerical
implementations (for example in LAPACK) exist and work well, in general. 
However, these algorithms only produce the eigenvalues for a given set of
parameters; changing the parameters produces a new set of eigenvalues with no
clarity on which of these eigenvalues belong together to say form a `mode'.
This would not be a major problem if the eigenvalues stay far apart and could be
traced separately; this is in general not the case because multiple roots are
inevitable for such systems. In the neighbourhood of such roots, the eigenvalue
trajectories can tangle and cross and the task of determining which eigenvalue
belongs to which trajectory can become quite difficult.

The spectral maps for a given system are indispensable 
in solving initial / boundary value problems involving that system. 
An understanding of the topography
specified by these maps as signified by the knowledge of critical points of the
map like saddles and branch points is crucial to a correct solution of such
problems. The absence of a reliable method to sort the eigenmodes correctly
has hampered a proper investigation of these problems; for example, in the
stability of fluid flows, very often only the least stable mode has been 
pursued instead of
a detailed study of the modal structure. The situation is described 
extensively in \cite{Suslov-2006}.

Crossing of the eigenvalue trajectories (equivalently collision of the
eigenvalues) is the main obstacle in a successful
sorting of the eigenvalues. There are a variety of crossings that have been
detailed in literature; we describe some of these below (we use $\alpha$ and
$\omega$ to denote the variable parameter and the eigenvalue respectively) -
\begin{enumerate}
\item False crossing C1. The trajectories, when plotted in the $\omega$ plane,
appear to cross i.e. there exist points in the $\omega$ plane such that 
$\omega_m (\alpha_l) = \omega_n(\alpha_k)$ with $\alpha_l \ne \alpha_k.$
These are false crossings. In stability problems, where the focus is often
on the least stable mode, crossings in imaginary parts i.e.
$Im[\omega_m(\alpha)] = Im[\omega_n(\alpha)]$ lead to ambiguous definition of
various modes. However, unless the real parts also coincide at the same 
$\alpha,$ these are obviously false crossings.
\item True crossing. When the above mentioned $\alpha_l = \alpha_k$,  we
have a true crossing and the eigenvalues are degenerate (\cite{Kato-1995}).
However, there are two cases to distinguish -
\begin{enumerate}
\item The non-defective case C2. A full complement of eigenvectors exists i.e
the geometric multiplicity of the repeated eigenvalue equals its algebraic
multiplicity.
\item The defective case C3. The geometric multiplicity is less than the
algebraic multiplicity. A full complement of independent eigenvectors does
not exist and generalised eigenvectors have to be used to obtain a full
complement. Two further subcases can be distinguished -
\begin{enumerate}
\item The analytic case C3A. The eigenvalues are analytic functions of the
parameter.
\item The non-analytic case C3B. The eigenvalues are non-analytic functions of
the parameter; in particular, they have branch points (BPs).
\end{enumerate}
\end{enumerate}
\item Avoided crossing C4. The eigenvalues come very close but do not actually
collide. Avoided crossings can be `broad' or `sharp'; the latter case is the
same as case C3B. Avoided crossings are of importance in  the study of chemical
reaction mechanisms and in quantum mechanics, especially to an understanding of
quantum chaos.
\end{enumerate}

Many approaches to sort eigenvalues exist in literature. We list some of these
below -
\begin{enumerate}
\item Ascending / descending order of their real or imaginary parts (S1).
\item Absolute value (S2).
\item Nearest neighbour (S3).
\item A combination of nearest neighbouring eigenvalue and nearest neighbouring
eigenvector (S4).
\end{enumerate}

S1 is the simplest to use and is reliable if the eigenvalue trajectories are
far apart. However, S1 fails even at false crossings C1. Examples are given
in \cite{Koch-1986} and \cite{Suslov-2006} which consider stability problems in 
fluid mechanics. S2 and S3 negotiate the false crossings most of the time but 
can fail sometimes (for example when the crossing point is on the real line) and
at true crossings. S4, which is used in some of the Open Source subroutines 
(eigenshuffle.m, eigenshuffle.py \cite{eig1,eig2}) is
succesful more widely but can fail at certain types of C3 crossings. We will
provide examples of these failures in Sec 3.

In this paper, we present a simple method to sort eigenmodes 
from eigenvalues that have been produced from an eigenvalue solver. The only
assumption is that the dispersion relation is an analytic function of its
parameters. \footnote{This is often the case for physical problems.} The
eigenvalues are sorted, based on their imaginary parts, at an initial point. 
From these, the neighbouring values of each mode are obtained by using the
Cauchy - Riemann equations. This works in general because the eigenmodes
are analytic 
and in particular at false collisions and true collisions of type C2 and C3A.  
Even though the C-R criterion cannot be employed at branch points (for e.g. 
a double root in the $\omega$ plane), the
algorithm can still educe such critical points. The mode-sorting algorithm
is explained in more detail in Sec. 2. We demonstrate the method on three
problems in Sec. 3. We use the algorithm to sort the temporal even 
Orr-Sommerfeld (OS) modes that arise in a stability analysis of 
Couette flow in 3.1. In 3.2, we apply the algorithm to sort the
jumbled eigenvalues in a model problem involving a cube-root branch point.
Finally, we sort the eigenpaths of two low dimensional
matrices in 3.3 and 3.4 and show how some currently available methods fail to 
produce the correct eigenpaths.
\section{Mode sorting} \label{sec:mode}
\subsection{The algorithm}
SPEC-RE (Sorting Procedure for Eigenvalues based on Cauchy - Riemann Equations)
is based on a well-known result of function theory for polynomials \footnote{
 The roots of such polynomials are then analytic functions of the same parameter
 {\sl{with only algebraic singularities}} (p.64, \cite{Kato-1995}.)}
whose coefficients are analytic functions of a parameter.
Thus, the eigenvalue $\omega_j$ is an analytic function
 of the complex wavenumber $\alpha$ except at isolated branch points.  At any
such point of analyticity $\alpha=\alpha_p$, the quantity
 $$\displaystyle  F=\left|\frac{\partial{\mathcal{R}e}(\omega_j)}{\partial
\alpha_r}-
\frac{\partial{\mathcal{I}m}(\omega_j)}{\partial\alpha_i} \right|+
 \left|\frac{\partial{\mathcal{R}e}(\omega_j)}{\partial\alpha_i}+\frac{\partial{
\mathcal{I}m}
(\omega_j)}{\partial\alpha_r} \right|$$ has to be negligible
 since Cauchy-Riemann conditions for analytic functions have to be satisfied. In
 what follows, this quantity $F(\omega_j;\ \alpha_p)$ is called the CR residue.

The primary task of SPEC-RE is to sort each eigenvalue of the spectrum
from a given initial point in the $\alpha$ plane by minimization of the CR
residue at the points of analyticity.
The algorithm makes use of the negation of the CR criterion at the branch
points rather than prescribing a `method' to identify branch points. The
primary feature of the algorithm, for analytic points, is described in 2.2; 
how the algorithm can educe non-analytic points like branch points is 
explained in 2.3 and 2.4.
\begin{figure}
\centering
\includegraphics[width=4in]{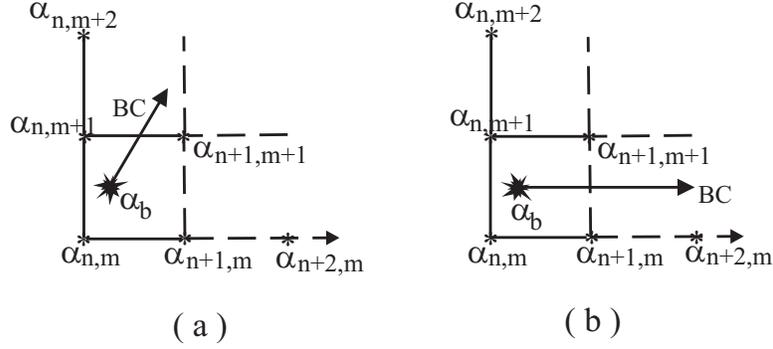}
\caption{Schematic to explain mode sorting around a BP $\alpha_b$ by the
algorithm SPEC-RE.}
\end{figure}
\subsection{Computation of analytic traces}
 The computational domain is a rectangular patch in the $\alpha$ plane with
edges parallel to the axes. The grid points are equally spaced along both the 
axes; however, the grid size in these directions may be different. The sorting
algorithm is implemented on a 4-point stencil of this grid (dashed line in
figure 1(a)); at any given pivot point $\alpha_{n,m}=(\alpha_r,\alpha_i)$
 the stencil consists of the neighbouring points $\alpha_{n-1,m} = (\alpha_r
- \delta \alpha_r, \alpha_i)$, $\alpha_{n+1,m}= (\alpha_r +
\delta \alpha_r,\alpha_i)$ and $\alpha_{n,m+1}=(\alpha_r,\alpha_i+\delta
\alpha_i)$. Given a particular eigenvalue
$\omega_j$ at $\alpha_{n,m}$, the algorithm is designed to pick one (and only
one) of the eigenvalues from the spectrum at two neighbouring points
$\alpha_{n+1,m}$ and $\alpha_{n,m+1}$
 such that the CR condition at $\alpha_{n,m}$ is satisfied. Equivalently, the
relevant complex derivatives at $\alpha_{n,m}$ must make the CR residue $F$
 to be negligible.  In the numerical procedure, these derivatives
are replaced by the central and forward differences
 $$\frac{\partial\omega_j}{\partial\alpha_r}=\frac{\omega_l(\alpha_{n+1,m})-
\omega_j(\alpha_{n-1,m})}{2 \delta\alpha_r},\   l=1,2,3,...$$
 $$\frac{\partial\omega_j}{\partial\alpha_i}=\frac{\omega_k(\alpha_{n,m+1})-
\omega_j(\alpha_{n,m})}{i\delta\alpha_i}, \  k=1,2,3,...$$
 \noindent The CR residue $F(\omega_j;\ \alpha_{n,m})$ is defined using these
central-forward differences and is actually a set of numbers
$F_{l
k}(\omega_j;\alpha_{n,m}) ; \ l=1,2,3,... \
 k=1,2,3,... $ .The indices $k_p$ and $l_p$  that correspond to the minimum of
these numbers for a given $j$, which is expected to be a negligible quantity,
are picked.  As the analyticity condition
  for $\omega_j$ at $\alpha_{n,m}$ is numerically satisfied between $\omega_j,
\omega_{l_p}$ and $\omega_{k_p}$, all three $\omega$s belong to the same
analytic function.
  In other words,
\begin{gather}
 \omega_j(\alpha_{n+1,m})=\omega_{l_p}, \tag{a}  \\
\omega_j(\alpha_{n,m+1})=\omega_{k_p}. \tag{b}
\end{gather}
  \noindent The pivot point can then be moved to one of the two adjacent points
either in the horizontal direction or the vertical
direction and the sorting procedure can be repeated for the new stencil.
Hence, starting from an initial point $\alpha_0$, the sorting procedure
picks one and only one value from the spectrum at each grid point and assigns
it to the $j$th collection
so that an analytic function $\omega_j(\alpha)$ is constructed, on the entire
rectangular patch in the $\alpha$ plane.
\subsection{Sweep direction}
In a horizontal sweep, the pivot point $\alpha_{n,m}$ moves along the direction
of increasing $\mathcal{R}e(\alpha)$, keeping $\mathcal{I}m(\alpha)$ constant.
After reaching the right-most point of the grid, the pivot point is moved to
 $\alpha_{1,m+1}$. Further computations are performed on stencils containing
$\alpha_{n,m+1}, \alpha_{n+1,m+1}$ and $\alpha_{n,m+2}$ starting from $n = 1.$
It may be noted that the eigenvalues at this level have already been sorted
from the computation at the $m$-th level, as shown in equation (b). Hence, 
using the eigenvalues at the $m + $1-st level, either (i) $\omega_j$ at the 
$m + 2$-nd level may be sorted, or (ii) re-sorting may be done afresh at the
$m+1$-st level.
Method (ii) will not produce any new arrangement\footnote{Re-sorting leads to a
re-arrangement of the eigenvalues, but, the re-arrangement may be the same as
the existing one.} of eigenvalues at the $m+1$-st level 
 unless a branch point lies between the $m$-th and the $m+1$-st
levels. Eduction of a branch cut along the sweep direction (horizontal) by
Method (ii) will be explained in the following subsection.

The sweep direction is not rigidly fixed. A vertical sweep, for instance, will
produce a different modal map, with vertical branch cuts. One could
indeed sweep even along any family of parametric curves; the C-R equations 
would then have to be
satisfied in the appropriate coordinates.
 
\subsection{Mode sorting around a branch point}
Assume that there exists a branch point between $\omega_j$ and $\omega_{j+k}$
located in the box formed by the $m$-th, $m+1$-st, $n$-th and $n+1$-st lines as shown in figure 1 (
i.e. $\omega_j(\alpha)$ and $\omega_{j+k}(\alpha)$ intersect at some
$\alpha_b$).
By design, the sorting algorithm produces an analytic $\omega_j$ not only
up to the $m$-th line, but also upto the point $\alpha_{n+1,m+1}$ on the 
$m+1$-st line.  At the stencil formed by $\alpha_{n+1,m} $, $\alpha_{n+2,m}
$ and $\alpha_{n+1,m+1} $,
application of CR condition forces analyticity of $\omega_j$
 at both edges of the stencil and hence, does not allow the BC to cut the
$\alpha_{n+1,m} $ - $\alpha_{n+1,m+1} $ edge. The forcing of analyticity on the
lower and left edges of the box by the previous stencil leads to the BC 
cutting the $\alpha_{n,m+1} $ - $\alpha_{n+1,m+1}$ edge, as shown
 in figure 1(a).
 If further computations were to be done using Method (i) to sort eigenvalues at
the $m+2$-nd level, application
of the CR condition for the stencil at $\alpha_{n,m+1}$ will be erroneous due
to the aforementioned  non-analyticity  at the $\alpha_{n,m+1} $ - $\alpha_{n
+1,m+1} $ edge. By Method (ii), $\omega_j$ values along
that line are rearranged and analytic sorting between $\omega_j(\alpha_{n,m+1})$
and $\omega_j(\alpha_{n+1,m+1})$ is ensured. Analyticity along this edge forces
non-analyticity of $\omega_j$ along the  $ \alpha_{n+1,m}$ -
$ \alpha_{n+1,m+1}$ edge, which is equivalent to the BC being horizontal in that
grid box as shown in figure 1(b). By continuation of the horizontal sweep at
the $m+1$-st level, a horizontal BC evolves naturally.
A vertical sweep, together with the application of
Method (ii) in the vertical direction would produce a vertical BC. 
It should, in principle, be possible to modify the algorithm to obtain a 
branch cut along a suitable complex curve from the branch point by allowing
non-analyticity at suitable edges of the stencils while sweeping. Such an
improvement will be useful in situations when some points in the $\alpha-$plane
are needed to be retained in an analytic region. However, it is not done here.
\section{Results \& Discussion}
We now present some results obtained using SPEC-RE in a few problems. We
start with a well-known problem from fluid mechanics.
\subsection{Sorting of modes in a physical problem}
The Orr - Sommerfeld (OS) equation is the basic equation of linear stability
analysis of parallel shear flows and is given by \cite{Schmid-2001}
\begin{gather}
\bigg{[} i \alpha U (D^2 - k^2) - i \alpha D^2 U - \frac{1}{Re} (D^2 - k^2)^2 
\bigg{]} v = i \omega (D^2 - k^2) v
\end{gather}
where $U(y)$ is the base streaming flow, $\alpha$ is the streamwise wavenumber,
$k^2 = \alpha^2 + \beta^2$ where $\beta$ is the spanwise wavenumber, $D$ is
differentiation wrt $y, Re $ is the Reynolds number and $\omega$ the frequency.
$v$ is the normal perturbation velocity. 
\begin{figure}
\centering
\includegraphics[width=5in]{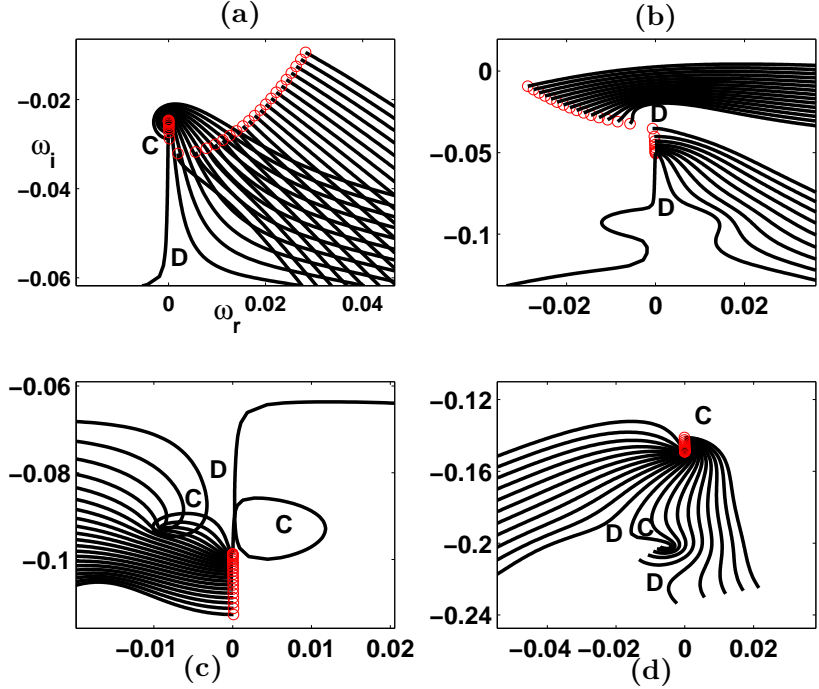}
\caption{Temporal maps (Example 3.1) of the region $\textit{R} = [0,0.5] \times
[0,0.5]$ in the $\alpha$ plane. a) mode 1 b) mode 2 c) mode 3 d) mode 4. 
The red circles indicate starting points of the trajectories. `C' and `D' 
indicate cusps and double roots respectively.}
\end{figure}

For a given $U(y)$, a Chebyshev discretisation of Eq.(1) results in a NDR 
which can then be solved for either $\alpha$ or $\omega.$ We solve for
We solve for $\omega$ and sort the temporal eigenvalues, into temporal 
eigenmodes for the Couette flow $U(y) = y, y \in [1-,1]. \beta = 0$ is assumed.
We show in figure 2, the maps of the region $ \textit{R} = [0,0.5] \times
[-0.5,0.5]$ (in the
$\alpha$ plane ) for the lowest four OS modes. The maps are analytic at
almost all points, the only exceptional points being those at which branching
occurs. It is easy to see that basing the sorting on the real or imaginary parts
of $\omega$ will fail. Even other methods suggested in \cite{Suslov-2006} 
fail, as detailed there (figure 1 of that paper), though this has not been 
shown here.

\subsection{A model problem with a higher order branch point}
 \begin{figure}
\centering
\includegraphics[width=4.0in]{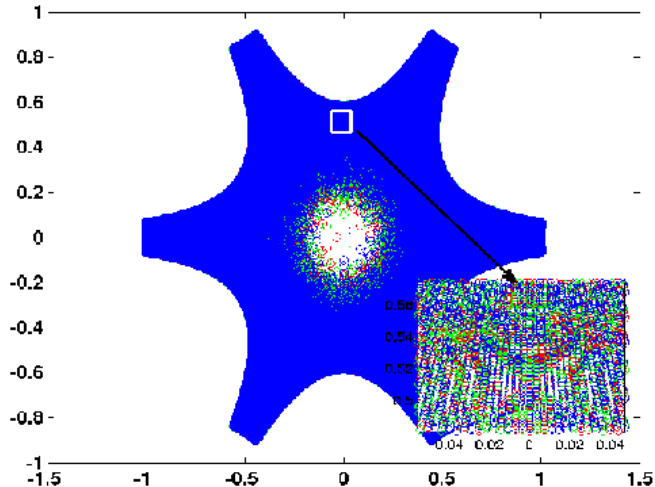}
\caption{Three branches of $f(z) = z^{1/3}$ presented in jumbled form.}
\end{figure}

\begin{figure}
\centering
\includegraphics[width=4.0in]{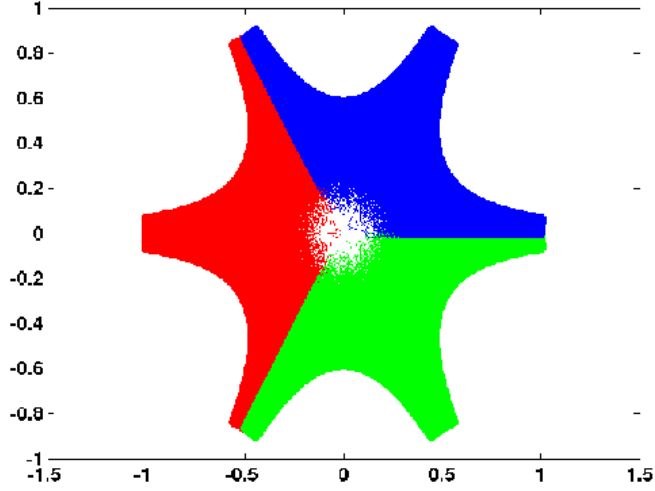}
\caption{SPEC-RE sorts the three branches correctly.}
\end{figure}
The previous example educed square-root branch points.
The sorting algorithm, in general, can also educe higher order branch points.
In this section, the applicability of SPEC-RE around a cube root branch point
is demonstrated. Here, the $\omega_j, \ \ j=1,2,3$ are the three branches of
the complex function (which can be thought as a factor in a
characteristic polynomial) $f(z)=z^{1/3}$. The grid size is chosen as 
$(\delta\alpha_r,\delta
\alpha_i) = (5e-4,5e-4)$ The three sets of $\omega$s are randomly jumbled at
every point in the $\alpha$ plane, as shown in
figure 3, and then read by the sorting program. The three sets of sorted 
values under a horizontal sweep are shown in different colors in figure 4.
\begin{figure}[h]
\centering
\subfigure[]{
\includegraphics[width=2.5in]{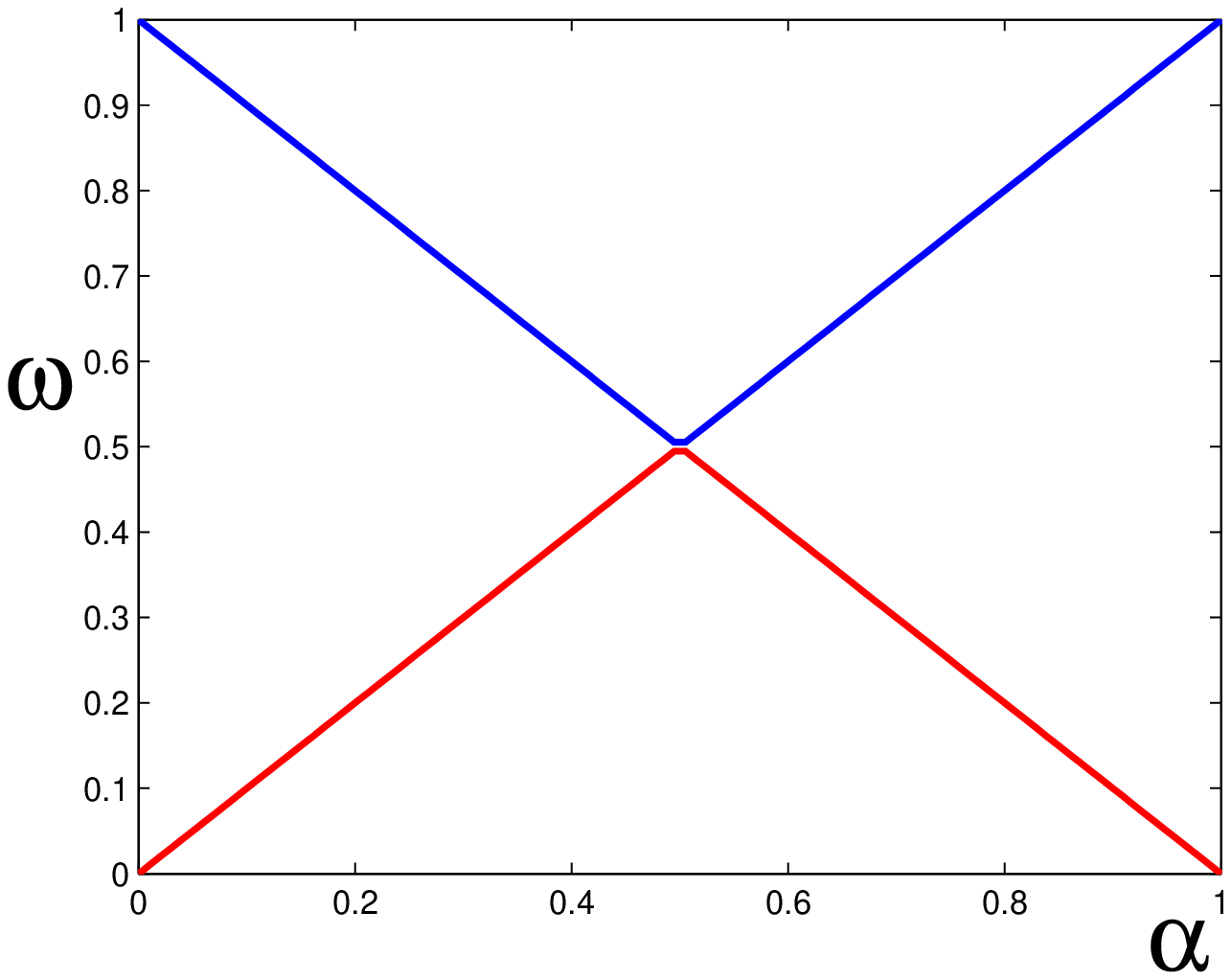}}
\hspace{0.1in}
\subfigure[]{
\includegraphics[width=2.5in]{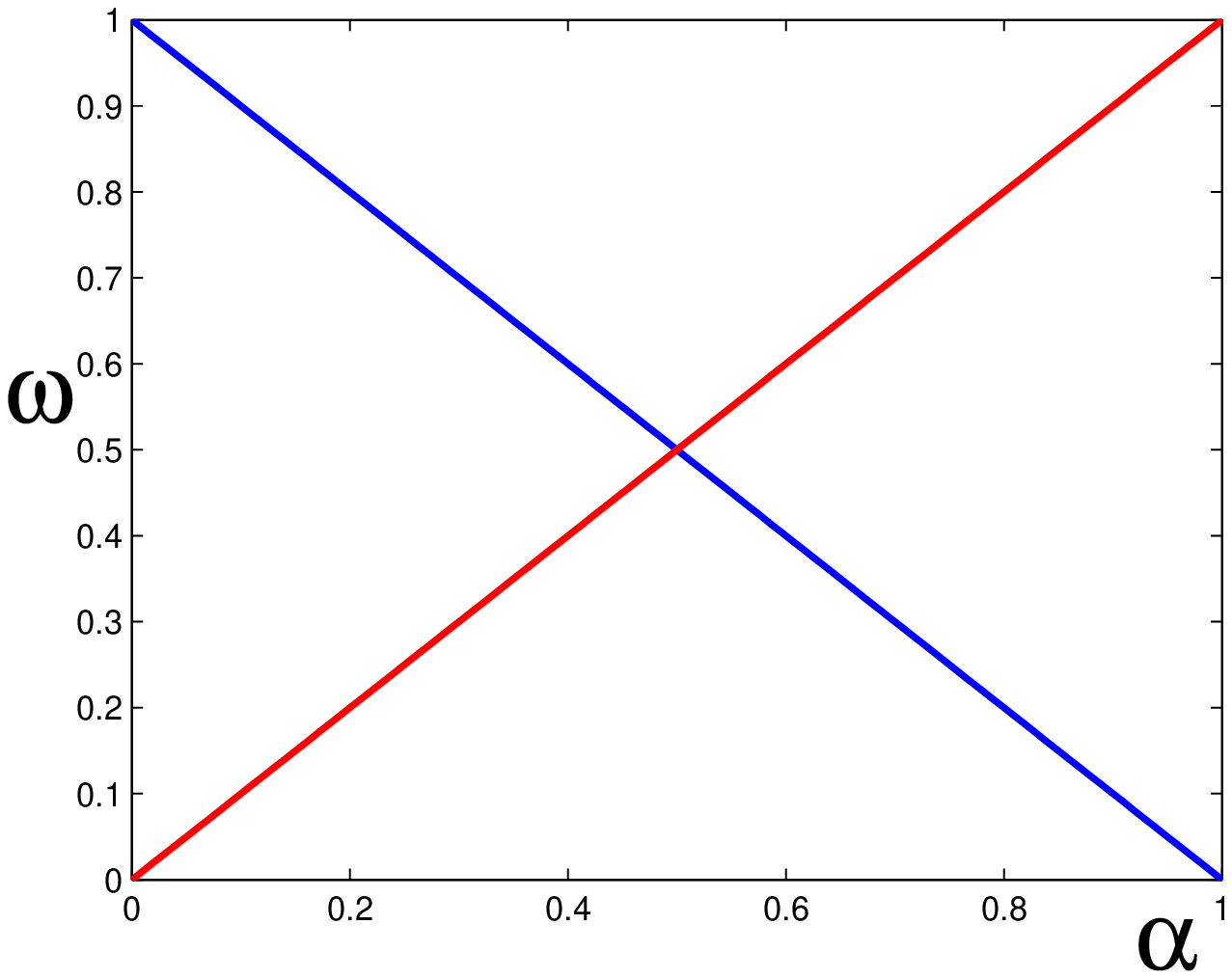} }
\caption{Eigenmodes produced by sorting schemes (a) $S_4$  and (b) SPEC-RE. 
$S_4$ interchanges the eigenvalues beyond the degenerate point $\alpha = 0.5$
whereas SPEC-RE sorts them correctly.}
\end{figure}

\subsection{$2 \times 2$ normal and non-normal matrices}
For the next example, we use two simple $2 \times 2$ matrices, given by
\begin{gather*}
A_1 = \begin{bmatrix}
\alpha & 0 \\
0 & 1 - \alpha \end{bmatrix}, \,\,
A_2 = \begin{bmatrix}
\alpha & 0 \\
\epsilon & 1 - \alpha \end{bmatrix},
\end{gather*}
where $\alpha \in [0,1]$ and $\epsilon$ is a fixed positive number. Note that
$A_1$ is normal whereas $A_2$ is non-normal; they have the same eigenvalues
$\omega_1 = \alpha$ and $\omega_2 = 1 - \alpha$ however. Sort $S_3$ fails to
produce $\omega_1$ and $\omega_2$ for both $A_1$ and $A_2$.
Sort $S_4$ produces the correct eigenpath for $A_1$ but fails for $A_2.$ 
SPEC-RE produces the correct
sort in both cases. The correct eigenpaths and the mis-sorted ones are shown
in figure 5, for $\epsilon = 0.1$.
$A_2$ has defective eigenvalues, and consequently a
generalised eigenvector, at $\alpha = 0.5$; $S_4$ which makes use of both
eigenvalue and eigenvector distance fails to sort correctly. In general, $S_4$  
is expected to fail in such defective situations. This leads us to our last
example involving a $5 \times 5$ matrix.
\subsection{A $5 \times 5$ matrix with defective eigenvalues}
Consider the $5 \times 5$ matrix whose non-zero elements are given by
\begin{gather*}
B(1,1) = \alpha (1 + i) + \alpha^2, \, B(2,2) = \alpha(- \cos \theta_1 + i 
\sin \theta_1) + \alpha^2, \,\, B(3,3) = \alpha (- \cos \theta_2 + i  
\sin \theta_2) + \alpha^3, \\
B(4,4) = \alpha (\cos \theta_3 - i \sin \theta_3) + \alpha^4, \,\, 
B(5,5) = \alpha ( \cos \theta_4 + i \sin \theta_4) + \alpha^5, \\
{\rm{and}} \,\,B(2,1) = B(3,2) = B(4,3) = 
B(5,4) = 0.1.
\end{gather*}
$\theta_1 = 3 \pi / 20, \theta_2 = 11 \pi / 20, \theta_3 = 7 \pi / 20,
\theta_4 = \pi / 20$ and $B(i,i)$ have been chosen to produce a quintuple
eigenvalue of 0 at $\alpha = 0.$ $\alpha \in [-1,1].$ The results of sorting by 
eigenshuffle and by the current algorithm 
SPEC-RE are shown in figure 6(a,b) respectively. The starting 
eigenvalues, for $\alpha = -1$ are marked with a star in fig 6(a) and the
end eigenvalues, for $\alpha = 1$ are marked with a circle in fig 6(b).
$S_4$ jumbles the eigenvalues at the defective point (fig 6(a)), so that the
green curve
continues as red, the blue as black, the black as magenta, the red as blue and
finally the magenta as green. SPEC-RE maintains the correct eigenvalue 
positions for the entire $\alpha$ range (fig 6(b)).
\begin{figure}[h]
\centering
\subfigure[]{
\includegraphics[width=2.5in]{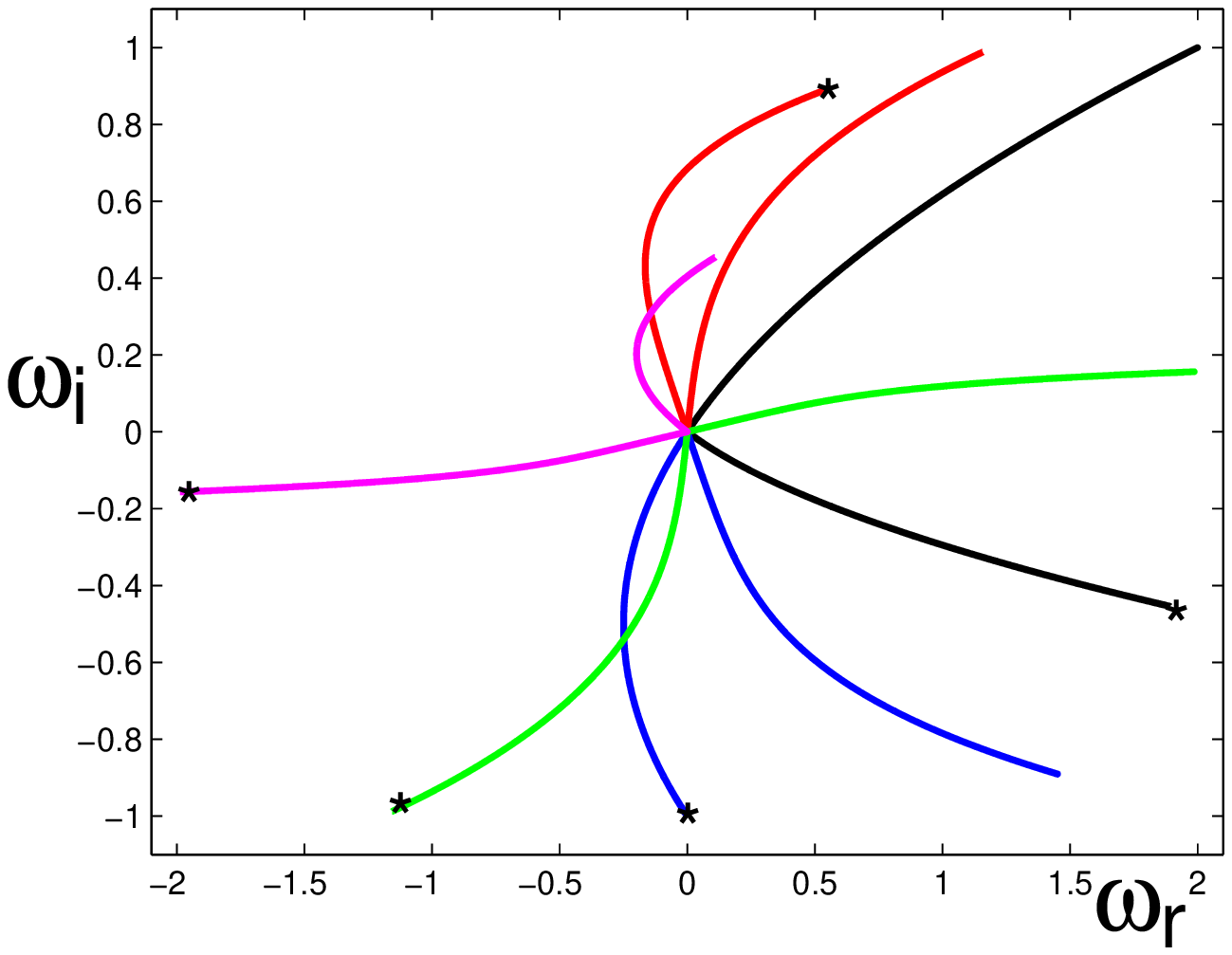}}
\hspace{0.1in}
\subfigure[]{
\includegraphics[width=2.5in]{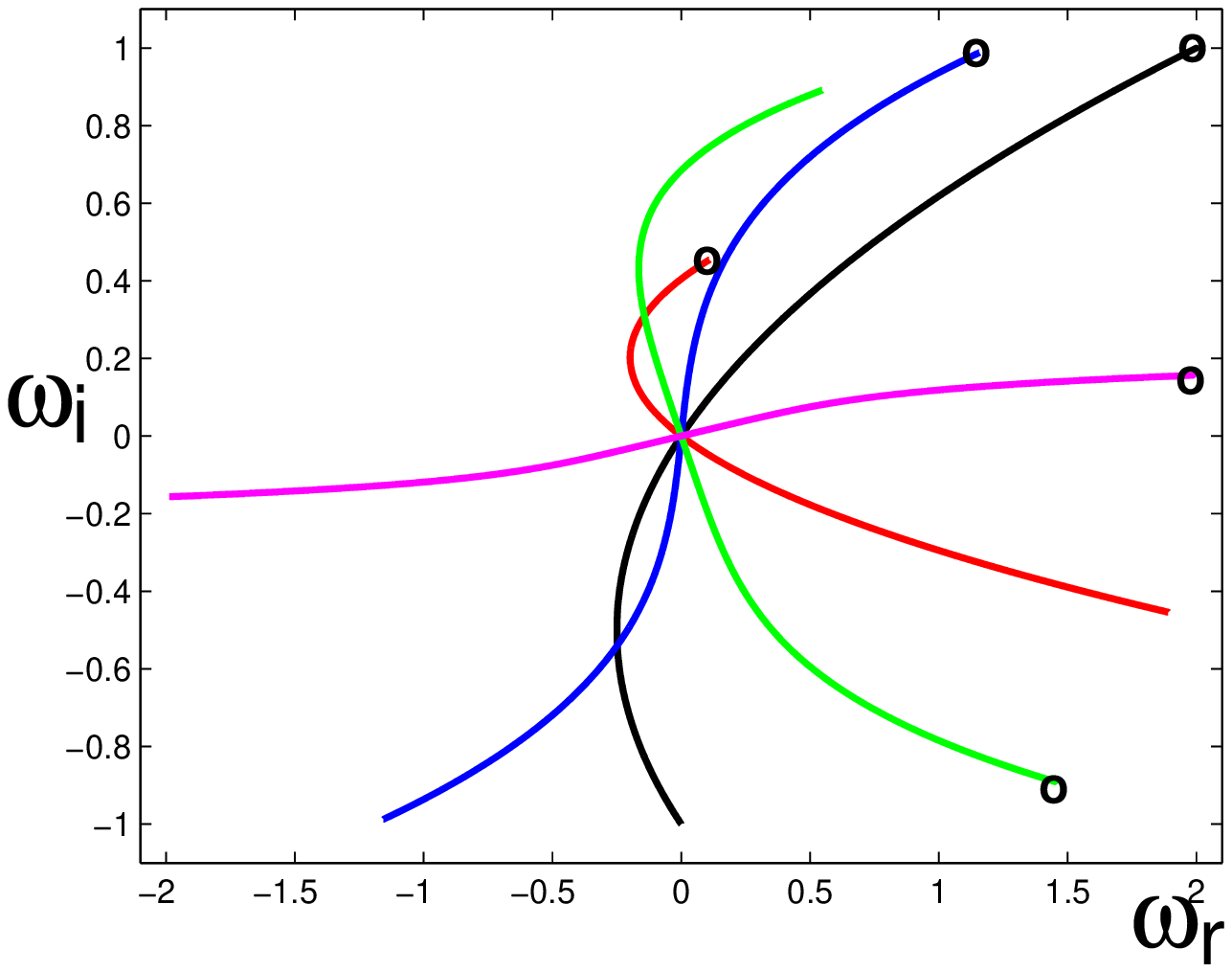} }
\caption{Eigenmodes of the $5 \times 5$ matrix $B$ of Example 3.4. The 
parameter $\alpha \in [-1,1].$ Sorting scheme $S_4$ produces the eigenpaths in
(a). Clearly, there is a mix-up of the modes beyond the quintuple root $\alpha
= 0.$ SPEC-RE produces the five correct eigenpaths in (b). The five eigenvalues 
at $\alpha = -1$ are marked
with a star in (a) and at $\alpha = 1$ by a circle in (b).}
\end{figure}

\begin{figure}
\centering
\includegraphics[width=3.0in]{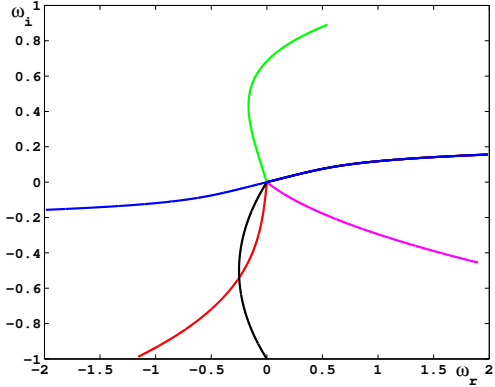}
\caption{Use of an L stencil in Example 3.4 results in missing eigenmodes 
beyond the quintuply degenerate eigenvalue at zero.}
\end{figure}
The last two examples show how true crossings can be misinterpreted as
avoided crossings (of the sharp kind) if the sorting is not proper. This 
could lead to a wrong inference of eigenstate exchange when none exists. 
In systems which allow both C3A and C3B type crossings, this could lead to
misidentification of eigenmodes leading to wrong conclusions.

For the application of this algorithm, the initial sorting point should be a 
point where all eigenvalues are simple. The C-R equations involve first order
derivatives in two directions; a numerical implementation could, in principle
use finite differences of any order.
For sorting at the point next to the initial one, we use a 3-point L stencil
(forward differences in both directions) and 4-point T stencils (central 
difference in the sweep direction and forward in the other) for the 
subsequent grid points. The T stencil is found to be necessary as the L
stencil can result in improper sorting under certain circumstances, for
example, if a grid point corresponds to a multiple eigenvalue. In the latter
case, the same mode can be picked multiple times since the mode is chosen
based on the forward difference, which involves the multiple eigenvalue, at
which grid point the modes are indistinguishable. A T stencil avoids this by 
including an upstream value as well, by using a central difference. Usage of
an L stencil in Example 3.4 results in the wrong eigenmodes shown in figure 7.

\section{Conclusion} \label{sec:conclusion}

An algorithm SPEC-RE, based on the analytic structure of the dispersion 
relation, has been devised to sort complex eigenvalues into modal
families. The method makes full use of the fact that the eigenvalues are 
analytic functions of the parameters, something that the methods currently in
existence do not.
Collisions of eigenvalues, whether false or true, are handled easily
unlike say in \cite{Suslov-2006}. The method can be easily extended to
multi-parameter mode sorting, with the C-R criterion being applied with respect
to one parameter at a time, with all other parameters kept fixed. Even 
generalised eigenvalue problems of polynomial type are amenable to this method 
because these can be transformed to linear
eigenvalue problems which standard routines can solve. This algorithm can be
even used to sort numbers that are not necessarily eigenvalues, as long as they
are obtained by an analytic process, for e.g. they could be roots of a 
transcendental equation. 

\section*{Acknowledgments}
The support of National Board of Higher Mathematics, India is gratefully  
acknowledged.


\end{document}